\documentclass{article}
\usepackage{spconf,amsmath,amssymb,amsfonts}
\usepackage{graphicx}
\usepackage{textcomp}
\usepackage[safe]{tipa}
\usepackage{booktabs}
\usepackage{enumitem}
\usepackage{url}
\usepackage[hidelinks]{hyperref}

\newcommand{\sys}{\textsc{EmpDPO}}
\usepackage{tikz}
\usetikzlibrary{positioning,arrows.meta,shadows,backgrounds,fit,calc}
\newcommand{\emptag}[1]{\texttt{\textless emp\textgreater}#1\texttt{\textless/emp\textgreater}}
\newcommand{\pairtag}[3]{{\setlength\fboxsep{1.5pt}\setlength\fboxrule{0.4pt}%
\fcolorbox{#1}{#2}{\scriptsize\vphantom{chosen rejected $z_w z_l$}#3}}}
\newcommand{\chosentag}{\pairtag{green!45!black}{green!12}{chosen $z_w$}}
\newcommand{\rejectedtag}{\pairtag{red!60!black}{red!10}{rejected $z_l$}}

\makeatletter
\let\spconf@thebibliography\thebibliography
\def\thebibliography#1{\spconf@thebibliography{#1}\setlength{\itemsep}{0pt}\setlength{\parsep}{0pt}}
\makeatother

\title{\sys: Enabling Emphasis Control in TTS via \\ Direct Preference Optimization}

\name{Haonan Chen$^{1}$, Yunyun Wang$^{1}$, Megan Wei$^{2}$, Zeyu Jin$^{1}$\thanks{Audio samples: {\urlstyle{same}\url{https://chaonan99.github.io/empdpo-demo/}}}}
\address{$^{1}$Adobe Research, USA \qquad $^{2}$Brown University, USA}

\makeatletter
\def\ps@arxivnotice{%
  \let\@oddhead\@empty\let\@evenhead\@empty
  \def\@oddfoot{\parbox[b]{\textwidth}{\footnotesize\raggedright This work has been submitted to the IEEE for possible publication. Copyright may be transferred without notice, after which this version may no longer be accessible.}}%
  \let\@evenfoot\@oddfoot}
\makeatother
\begin{document}
\ninept
\raggedbottom
\setlength{\textfloatsep}{12pt plus 2pt minus 2pt}
\setlength{\floatsep}{8pt plus 2pt minus 2pt}
\setlength{\intextsep}{8pt plus 2pt minus 2pt}
\setlength{\dbltextfloatsep}{10pt plus 2pt minus 2pt}
\setlength{\dblfloatsep}{8pt plus 2pt minus 2pt}
\setlength{\abovecaptionskip}{5pt}
\setlength{\belowcaptionskip}{0pt}
\makeatletter
\renewcommand\section{\@startsection{section}{1}{\z@}{-2.6ex plus -.8ex minus -.2ex}{1.6ex plus .2ex}{\reset@font\large\bfseries\centering}}
\renewcommand\subsection{\@startsection{subsection}{2}{\z@}{-1.9ex plus -.6ex minus -.2ex}{1.1ex plus .2ex}{\reset@font\normalsize\bfseries}}
\makeatother
\maketitle
\thispagestyle{arxivnotice}

\begin{abstract}
Word-level emphasis conveys meaning that text alone cannot express and is therefore essential for a practical text-to-speech (TTS) system. In this work, we study how to add emphasis control post hoc to a pretrained TTS model that has never been trained with emphasis annotations. We present \sys, a simple, efficient, and effective two-stage recipe that adds emphasis control to a multilingual zero-shot diffusion TTS model through supervised fine-tuning followed by direct preference optimization (DPO). The training signals for both stages come from an off-the-shelf multilingual emphasis detector, so that no manual emphasis annotation or task-specific reward design is needed. We find that DPO can substantially strengthen emphasis control with only a few thousand preference pairs and training steps. Objective and subjective evaluations against strong public systems show that our model produces perceptible emphasis while preserving naturalness and overall quality.
\end{abstract}


\section{Introduction}
Current text-to-speech (TTS) systems can already generate highly realistic and natural speech \cite{chen2024f5tts,xie2025controllablesurvey}.
However, human speech carries information that is difficult to fully convey through text alone.
Emphasis is one such example: emphasizing different words in the same sentence leads to different contrastive focus \cite{deseyssel2024emphassess}.
For example, consider the sentence ``She called today.''
\begin{quote}
\emph{She} called today (not someone else).\\
She \emph{called} today (not messaged).\\
She called \emph{today} (not yesterday).
\end{quote}
Therefore, when TTS systems are used in practice, users may need finer-grained control to accurately convey the intended meaning.

A straightforward approach to fine-grained control is to treat control signals as additional input and learn their realization through supervised training.
For emphasis control, there are two groups of methods: (1) directly learning emphasis from emphasis tags or features fed to the model \cite{suni2020prominence,liu2021emphatic,shechtman2021lexicalfocus,guo2022prosodytagging,zhong2023eetts}; and (2) controlling emphasis indirectly through more concrete prosodic attributes, such as pitch, energy, and duration \cite{seshadri2022emphasis,bauer2025explicit,joly2023zerodata,nie2026wordvoice}.
The former leaves the acoustic realization of emphasis entirely to the model, but requires training data with word-level emphasis annotations, which are costly to obtain; the latter restricts emphasis to a few prosodic parameters and tends to be language-dependent and less natural than end-to-end generation.

To further improve the adherence to control signals, recent work has explored feedback-based post-training for fine-grained speech control, ranging from online reinforcement learning such as Group Relative Policy Optimization (GRPO) \cite{shao2024deepseekmath} to offline methods such as Direct Preference Optimization (DPO) \cite{rafailov2023dpo,wallace2024diffusiondpo}.
For example, EMORL-TTS \cite{li2026emorltts} defines emphasis as a pitch/energy contrast reward computed on generated speech and reinforces it with a GRPO-style objective.
Fish Audio S2 \cite{liao2026fishs2} first uses supervised fine-tuning (SFT) to teach localized vocal instructions such as \texttt{[emphasis]}, and then applies GRPO to generated speech, using a rich transcriber to penalize omitted instructions.

Inspired by these approaches, we adopt a two-stage recipe---SFT followed by preference optimization---to add multilingual emphasis control to a pretrained TTS system that has never seen emphasis labels.
Our method involves three key design choices.
First, for preference optimization, we use DPO, which learns directly from fixed chosen/rejected pairs without requiring a separate reward model.
Second, we use the same off-the-shelf multilingual emphasis detector \cite{wei2026emphasisgeneralize} to generate pseudo-labels for both SFT and preference optimization, eliminating the need for manually annotated emphasis data.
Third, we build on a latent diffusion TTS backbone trained with flow matching \cite{lipman2023flowmatching}, similar to the teacher model of DMOSpeech \cite{li2024dmospeech}, which provides strong zero-shot voice cloning and efficient inference.
Our main contributions are as follows:
\begin{itemize}[leftmargin=0pt,labelwidth=0pt,labelsep=0pt,itemindent=0pt,nosep,label={},itemsep=2pt]
\item \textbf{A simple recipe.} We present a detector-in-the-loop SFT and Diffusion-DPO recipe for emphasis control that needs no manual emphasis annotation or task-specific reward design.
\item \textbf{Efficient training.} We show that DPO substantially strengthens emphasis control with only a few thousand preference pairs and training steps, and that the gain holds across six languages.
\item \textbf{Effective control.} Objective and subjective evaluations against strong public systems show that our model produces perceptible emphasis while preserving naturalness and overall quality.
\end{itemize}

\begin{figure*}[t]
\centering
\resizebox{0.9\textwidth}{!}{\begin{tikzpicture}[
  font=\scriptsize,
  >={Stealth[length=2mm,width=1.6mm]},
  node distance=4mm and 5mm,
  every node/.style={align=center},
  flow/.style={->, line width=0.6pt, draw=black!65},
  dflow/.style={->, line width=0.6pt, draw=black!55, dashed},
  box/.style={draw=black!45, fill=white, rounded corners=2pt, inner xsep=4pt, inner ysep=3pt,
              minimum height=8mm,
              drop shadow={opacity=0.12, shadow xshift=0.5pt, shadow yshift=-0.7pt}},
  model/.style={box, draw=blue!55!black, fill=blue!12, line width=0.8pt},
  det/.style={box, draw=orange!85!black, fill=orange!25, line width=0.8pt},
  loss/.style={box, draw=green!45!black, fill=green!13, line width=0.8pt},
  panel/.style={fill=black!5, rounded corners=5pt, inner sep=5pt},
  ptitle/.style={anchor=north west, font=\scriptsize\bfseries, text=black!75, inner sep=2pt},
  lbl/.style={font=\fontsize{6}{7}\selectfont, text=black!65, inner sep=1pt},
  mlbl/.style={font=\scriptsize, text=black!65, inner sep=1pt},
]
\node[box, anchor=west] (corpus) at (0,0) {speech corpora\\{\fontsize{6}{7}\selectfont waveform + transcript}};
\node[det, right=of corpus] (det1) {multilingual\\emphasis detector};
\node[box, right=of det1, minimum width=20mm] (scores)
  {per-word scores\\[9pt]{\fontsize{6}{7}\selectfont She\hspace{2.2mm}called\hspace{2mm}today}};
\node[box, right=9mm of scores] (marked)
  {She \emptag{called} today\\ \textipa{Si:} \emptag{\textipa{kO:ld}} \textipa{t@deI}};
\node[model, right=of marked] (base) {base diffusion\\TTS};
\node[model, right=10mm of base] (sft1) {SFT model};

\draw[flow] (corpus) -- (det1);
\draw[flow] (det1) -- (scores);
\draw[flow] (scores) -- node[lbl, above]{threshold} (marked);
\draw[flow] (marked) -- (base);
\draw[flow] (base) -- node[lbl, above]{fine-tune} (sft1);

\begin{scope}[shift={($(scores.center)+(-0.62,-0.20)$)}]
  \fill[black!30]          (0.02,0) rectangle (0.24,0.10);
  \fill[orange!85!black]   (0.50,0) rectangle (0.72,0.36);
  \fill[black!30]          (1.00,0) rectangle (1.22,0.15);
  \draw[densely dashed, red!70!black, line width=0.4pt] (-0.03,0.24) -- (1.27,0.24);
\end{scope}

\node[box, anchor=west] (inp) at (0,-2.35) {text\\+ reference audio};
\node[model, right=of inp] (gen) {SFT model\\{\fontsize{6}{7}\selectfont (sampling)}};
\node[box, minimum width=17mm, anchor=west] (cands) at ([xshift=5mm,yshift=4.6mm]gen.east) {$K$ marked\\candidates};
\node[box, minimum width=17mm, anchor=west] (plain) at ([xshift=5mm,yshift=-4.6mm]gen.east) {$N$ plain\\baselines};
\node[det, anchor=west] (det2) at ([xshift=6mm]cands.east |- gen.center) {multilingual\\emphasis detector};
\node[box, right=of det2] (pair)
  {pair selection\\[0.5pt]$\delta_k,\ \rho_k \;\rightarrow\; g_k$\\[2pt]\chosentag\;\rejectedtag};
\node[loss, right=of pair] (dpo) {Diffusion-DPO loss\\+ replay denoising loss};
\node[model, right=of dpo] (final) {\sys};

\draw[flow] (inp) -- (gen);
\draw[flow] (gen.east) -- ++(0.2,0) |- (cands.west);
\draw[flow] (gen.east) -- ++(0.2,0) |- (plain.west);
\draw[flow] (cands.east) -- ++(0.2,0) |- (det2.west);
\draw[flow] (plain.east) -- ++(0.2,0) |- (det2.west);
\draw[flow] (det2) -- (pair);
\draw[flow] (pair) -- (dpo);
\draw[flow] (dpo) -- (final);

\draw[flow] (sft1.south) -- ++(0,-0.35) -| ([xshift=3mm]gen.north);

\draw[dflow, <->, draw=orange!85!black] ([xshift=6mm]det1.south) to[out=-70, in=110, looseness=0.7]
  node[lbl, fill=white, text=orange!80!black, inner sep=1pt, pos=0.68, right=2pt]{same detector} ([xshift=2mm]det2.north);

\coordinate (t1) at ([yshift=3.6mm]det1.north);
\coordinate (t2) at ([yshift=3.6mm]cands.north);
\begin{scope}[on background layer]
  \node[panel, fit=(corpus)(sft1)(scores)(marked)(t1)(final.east |- corpus)] (p1) {};
  \node[panel, fit=(inp)(final)(cands)(plain)(t2)(sft1.east |- inp)] (p2) {};
  \foreach \n in {cands,plain}{
    \draw[black!40, fill=white, rounded corners=2pt]
      ([shift={(1.4mm,1.4mm)}]\n.south west) rectangle ([shift={(1.4mm,1.4mm)}]\n.north east);
  }
\end{scope}
\node[ptitle] at (p1.north west) {SFT stage};
\node[ptitle] at ([xshift=0mm]p2.north west) {DPO stage};
\end{tikzpicture}}
\vspace{-0.5em}
\caption{Overview of \sys. SFT stage: a multilingual emphasis detector labels real speech with emphasis markers. DPO stage: the same detector scores candidates sampled from the SFT model to build chosen/rejected pairs for Diffusion-DPO.}
\label{fig:overview}
\end{figure*}
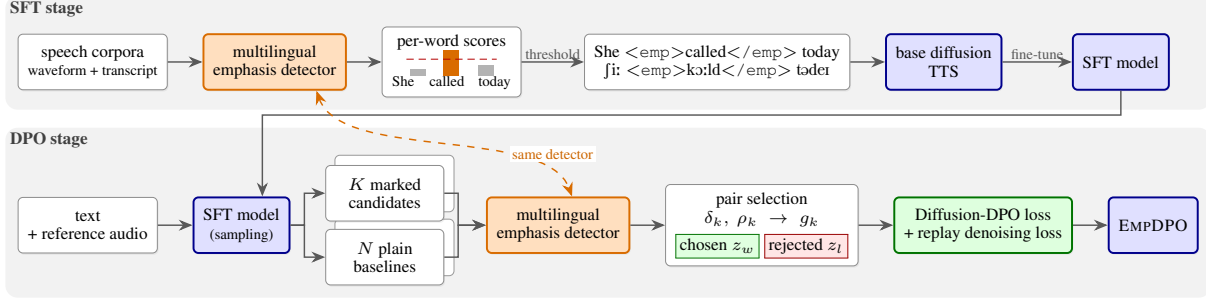

\section{Method}
\subsection{Base model}
Our base model is a multilingual zero-shot latent diffusion TTS model, similar to the teacher model of DMOSpeech \cite{li2024dmospeech}.
Its input consists of three parts: the raw text $x$, an IPA phoneme sequence $p$, and a reference audio $r$.
The text is encoded by a frozen pretrained T5 encoder into $e_x = E_{\mathrm{T5}}(x)$ \cite{raffel2020t5}, and the phonemes by a learned embedding table into $e_p = E_{\mathrm{ipa}}(p)$.
Both the target audio $y$ and the reference audio $r$ are encoded by the same frozen Descript Audio Codec (DAC) \cite{kumar2023dac} into latent frames $z = E_{\mathrm{DAC}}(y)$ and $z^r = E_{\mathrm{DAC}}(r)$.
The generation process can be written as a single conditional distribution:
\begin{equation}
\hat{z} \sim \pi_\theta(\cdot \mid e_x,\, e_p,\, z^r)
\end{equation}
and DAC then decodes the result back to a waveform $\hat{y} = D_{\mathrm{DAC}}(\hat{z})$.

The diffusion model is trained to denoise Gaussian noise into the latent $z$.
Specifically, $z$ is interpolated with noise $\epsilon$ to obtain a noisy latent $z_t = (1-t)z + t\epsilon$, where $t \in [0,1]$ is the noise level.
The model predicts a velocity $v_\theta$, which is trained to match the true velocity $v$ \cite{salimans2022progressive,lipman2023flowmatching}:
\begin{equation}
\label{eq:denoise}
\mathcal{L}(\theta) = \mathbb{E}_{t,\, z,\, \epsilon}\left\Vert
v_\theta(z_t,\, t,\, e_x,\, e_p,\, z^r) - v \right\Vert^2
\end{equation}
where $v = \mathrm{d} z_t/\mathrm{d} t = \epsilon - z$.
At inference, we start from pure noise and numerically integrate the predicted velocity (e.g., with Euler steps) to obtain the clean latent $\hat{z}$.

\subsection{Supervised fine-tuning with emphasis}
\label{ssec:sft}
To add emphasis control on top of the base model, we wrap each emphasized word with markers on both the text and phoneme paths.
For example,

\par\vspace{2pt}
{\centering
Text input: She \texttt{\textless emp\textgreater}called\texttt{\textless/emp\textgreater} today.\\
Phoneme input: \textipa{Si:} \texttt{\textless emp\textgreater}\textipa{kO:ld}\texttt{\textless/emp\textgreater} \textipa{t@deI}\par}
\vspace{2pt}\noindent We use an off-the-shelf multilingual emphasis detector \cite{wei2026emphasisgeneralize}.
The detector is trained on seven languages (en, de, es, fr, it, pt, zh) and is applied zero-shot to other languages.
Given a waveform and its transcript, it assigns an emphasis score to each word.
We further pick a threshold (0.35) and mark every word whose score exceeds it, on both the text and phoneme paths.
The model is then fine-tuned on a mixture of marked and unmarked data, giving it an initial ability to follow emphasis markers.

\subsection{Direct preference optimization (DPO)}
\label{ssec:dpo}
After SFT, we further strengthen the emphasis-following ability with DPO.
DPO training requires a pair of \emph{chosen} and \emph{rejected} targets for the same input, where the chosen one is preferred.
For emphasis control, the chosen audio should sound more emphasized than the rejected one.
Both are encoded by the same DAC encoder into a chosen latent $z_w$ and a rejected latent $z_l$ (winner/loser).
We denote the conditioning input by $c = (e_x, e_p, z^r)$.
The original DPO objective raises the likelihood of the chosen latent relative to a reference model and lowers that of the rejected one:
\begin{equation}
\mathcal{L}_{\mathrm{DPO}}(\theta) = -\mathbb{E}_{(z_w,\, z_l)}\,
\log \sigma\!\big(\beta \,[\, r_\theta(z_w) - r_\theta(z_l) \,]\big)
\end{equation}
where $r_\theta(z) = \log \big(\pi_\theta(z \mid c) / \pi_{\mathrm{ref}}(z \mid c)\big)$ is the log-likelihood ratio against the reference model $\pi_{\mathrm{ref}}$, the final model of the SFT stage, which serves as the anchor.
$\beta$ controls the allowed deviation, and $\sigma$ is the sigmoid function.
Unlike online reinforcement learning, DPO requires only offline preference pairs, with no separate reward model.

For a diffusion model, the log-likelihood cannot be evaluated directly, so Diffusion-DPO \cite{wallace2024diffusiondpo} replaces it with the denoising objective and applies the same relative comparison at each noise level:
\begin{equation}
\mathcal{L}_{\mathrm{diff}}(\theta) = -\mathbb{E}\, \log \sigma\!\left(
-\beta \left[ \left( E_\theta^{w} - E_{\mathrm{ref}}^{w} \right) -
\left( E_\theta^{l} - E_{\mathrm{ref}}^{l} \right) \right]\right)
\end{equation}
$E_\theta = \left\Vert v_\theta(z_t,\, t,\, c) - v \right\Vert^2$ is the velocity-prediction error on the corresponding sample.
Intuitively, this asks the policy to drive the denoising error on chosen below that of the frozen reference model, and to push the error on rejected above it.

In practice, optimizing the preference loss alone is unstable: the model drifts away from the reference and the denoising quality of both chosen and rejected samples degrades.
We therefore add a replay term, i.e., the original denoising loss of Eq.~(\ref{eq:denoise}) on additional data, and optimize the total loss $\mathcal{L} = \mathcal{L}_{\mathrm{replay}} + \lambda\,\mathcal{L}_{\mathrm{diff}}$ with $\lambda = 0.1$.
The replay data can come from two sources: the real speech used for SFT (\emph{corpus replay}), or the chosen samples of the preference pairs themselves (\emph{chosen replay}).
Corpus replay anchors the model to the real data distribution, whereas chosen replay requires no additional data pipeline; we compare both in Sec.~\ref{ssec:replay}.

\subsection{DPO pair construction}
\label{ssec:pairs}
Given input text $x$ and reference audio $r$ (markers are added to $x$ and $p$ identically; we write only $x$ below), we use the SFT model to generate $K$ candidates whose text carries emphasis markers, and $N$ baselines without markers.
We score all candidates and baselines with the emphasis detector.

Denote the set of marked target words in $x$ by $\mathcal{T}$; the improvement of a target word $\tau\in\mathcal{T}$ in candidate $k$ and the selectivity of candidate $k$ are defined as
\begin{equation}
\delta_k(\tau)=s_k(\tau)-\bar{s}^0(\tau),
\quad
\rho_k = \frac{\big|\{ u \notin \mathcal{T} : s_k(u) < s_k^{\min} \}\big|}{\big|\{ u \notin \mathcal{T} \}\big|}
\end{equation}
where $s_k(\tau)$ is the detector score of target word $\tau$ in candidate $k$, $\bar{s}^0(\tau)$ is its average score over the $N$ baseline samples, and the selectivity $\rho_k$ is the fraction of non-target words whose scores fall below the lowest target-word score $s_k^{\min}=\min_{\tau\in\mathcal{T}} s_k(\tau)$.
The composite score of candidate $k$ is:
\begin{equation}
g_k = \lambda_{\mathrm{min}}\, \delta^{\mathrm{min}}_k +
\lambda_{\mathrm{avg}}\, \bar{\delta}_k + \lambda_{\mathrm{sel}}\, \rho_k
\end{equation}
where $\delta^{\mathrm{min}}_k = \min_{\tau\in \mathcal{T}} \delta_k(\tau)$ is the smallest improvement across target words, $\bar{\delta}_k = \mathrm{avg}_{\tau \in \mathcal{T}}\, \delta_k(\tau)$ is the average improvement, and $\lambda_{\mathrm{min}}$, $\lambda_{\mathrm{avg}}$, $\lambda_{\mathrm{sel}}$ are the weights of the three terms. We use $(\lambda_{\mathrm{min}}, \lambda_{\mathrm{avg}}, \lambda_{\mathrm{sel}})=(0.6,0.4,0.3)$.
Intuitively, the first two terms reward candidates that emphasize every target word (the minimum term penalizes leaving any of them out), while the selectivity term rewards candidates in which non-target words stay unemphasized, so the emphasis is not simply achieved by making the whole utterance louder.
When $|\mathcal{T}|=1$, i.e., with a single emphasized target word, $\delta^{\mathrm{min}}_k=\bar{\delta}_k=\delta_k$, and $\rho_k$ reduces to the fraction of non-target words scoring below the target word.
We take the candidates with the highest and lowest $g_k$ as chosen $z_w$ and rejected $z_l$, respectively.

\section{Experimental setup}

\subsection{Training configuration}
\label{ssec:train}
The model is trained in three stages: pre-training without emphasis markers, SFT on the marked/unmarked mixture, and DPO with preference pairs and replay.
During pre-training, the model is trained for 700k steps on 300k hours of internal speech data across 21 languages, including the six evaluation languages.
SFT uses about 258k high-quality training samples (711 hours), 45\% of which carry \texttt{\textless emp\textgreater} markers.

We then use the SFT model to generate the preference pairs.
For each input, we generate $K=5$ candidates and $N=3$ baselines.
In addition to the composite ranking score defined in Sec.~\ref{ssec:pairs}, we apply the following filters:
\begin{enumerate}[leftmargin=*,nosep]
\item The candidate must satisfy $\delta_k^{\min} \geq 0.20$
  ($\geq 0.15$ for ja/zh), ensuring that it realizes the intended emphasis to some degree.
\item Both chosen and rejected outputs must satisfy CER $\leq 0.20$, NISQA $\geq 2.0$,
  SIM $\geq 0.15$, and a clipping fraction $\leq 0.001$ to ensure speech
  quality.
\end{enumerate}
Here, CER is the character error rate between the Qwen3-ASR \cite{shi2026qwen3asr} transcript and the reference text; NISQA is the MOS predicted by a no-reference speech-quality model \cite{mittag2021nisqa}; and SIM is the cosine similarity between the speaker embeddings of the generated audio and the reference audio, computed by an ECAPA-TDNN speaker-verification model on WavLM-Large features \cite{chen2022wavlm,desplanques2020ecapa}.
Only pairs satisfying all of the above hard filters are included in the DPO dataset.
The resulting dataset contains 5{,}171 preference pairs (en 957, de 878, es 1{,}314, fr 796, ja 662, zh 564).
We run DPO for 10k steps on 8 GPUs with a batch of 32 pairs plus 64 replay clips per step, using AdamW with a learning rate of $2\times10^{-5}$ and $\beta=0.1$.

\subsection{Baselines}
\label{ssec:baselines}
\begin{table*}[t]
\caption{Objective evaluation results (210 sentences $\times$ 8 voices = 1,680
inputs per system; 560 for WordVoice).
}
\label{tab:objective}
\centering
\small
\begin{tabular}{lccccc}
\toprule
System & $\bar{\delta}\uparrow$ & CER plain\,/\,emph.\ $\downarrow$ & SIM plain\,/\,emph.\ $\uparrow$ & UTMOS plain\,/\,emph.\ $\uparrow$ & NISQA plain\,/\,emph.\ $\uparrow$\\
\midrule
Ours-Base & --- & \textbf{0.008}\,/\,--- & 0.49\,/\,--- & 3.61\,/\,--- & 4.34\,/\,---\\
Ours-SFT & 0.069 $\pm$ 0.003 & \textbf{0.008}\,/\,\textbf{0.007} & 0.47\,/\,0.47 & 3.60\,/\,3.58 & 4.42\,/\,4.42\\
Ours-DPO & \textbf{0.136 $\pm$ 0.005} & \textbf{0.008}\,/\,0.008 & 0.49\,/\,0.49 & 3.60\,/\,3.55 & 4.36\,/\,4.34\\
Fish-S2 & 0.129 $\pm$ 0.005 & 0.039\,/\,0.034 & \textbf{0.51}\,/\,\textbf{0.51} & \textbf{4.20}\,/\,\textbf{4.20} & 4.42\,/\,4.39\\
Qwen3-TTS & 0.017 $\pm$ 0.003 & 0.034\,/\,0.036 & N/A & 3.56\,/\,3.48 & 4.18\,/\,4.06\\
WordVoice (en/zh) & 0.074 $\pm$ 0.008 & 0.047\,/\,0.040 & 0.46\,/\,0.46 & 3.92\,/\,3.86 & \textbf{4.77}\,/\,\textbf{4.74}\\
\bottomrule
\end{tabular}
\end{table*}
We compare our approach against three public systems.

\noindent\textbf{Fish-S2}~\cite{liao2026fishs2} (Fish Audio S2 Pro) supports zero-shot voice cloning from a reference audio and provides an inline \texttt{[emphasis]} tag, which makes it the closest match to our setting.

\noindent\textbf{WordVoice}~\cite{nie2026wordvoice} introduces discrete word-level (character-level for Chinese) attribute tokens for duration, energy, pitch, boundary, and tone.
Attributes not specified are predicted by the model itself.
It has no explicit notion of emphasis and supports only English and Chinese.
To use it, we first perform a generation pass and do force-alignment on the output to obtain the timing of each word.
We then generate a second time with the target word's pitch set to the mean pitch of the input reference audio $+4.5$ semitones, its energy set to the mean $+4$\,dB, and its duration set to $1.45\times$ ($1.35\times$ for Chinese) the duration of that word in the first pass, to simulate emphasis.
These values were chosen in a small-scale listening test to balance emphasis strength and naturalness.

\noindent\textbf{Qwen3-TTS}~\cite{hu2026qwen3tts} (CustomVoice variant) controls emphasis through a natural-language \texttt{instruct} field but does not support voice cloning.
We therefore instruct the model to emphasize the target word and use its built-in voices.

\subsection{Evaluation protocol and metrics}
\label{ssec:eval}
\begin{figure}[t]
\centering
\includegraphics[width=0.75\columnwidth]{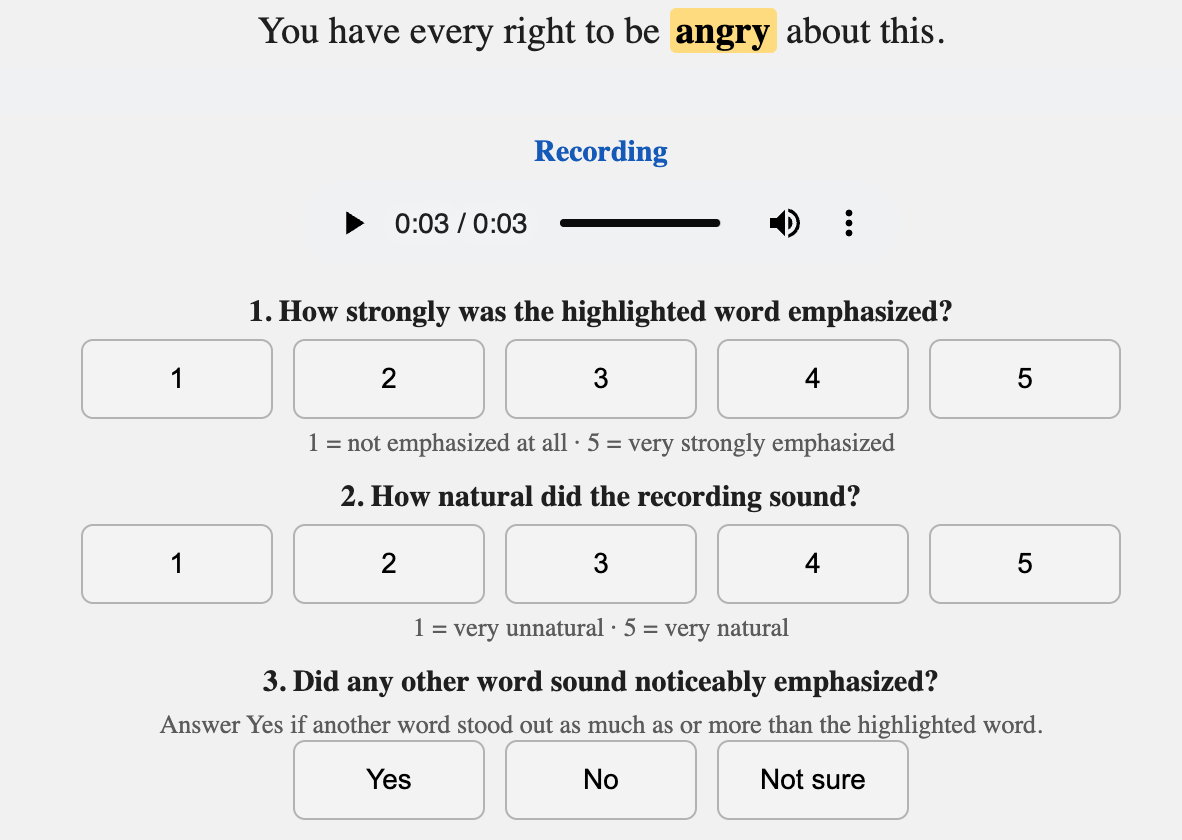}
\vspace{-0.5em}
\caption{Listening test interface (English version): emphasis strength of the highlighted word (Q1) and naturalness (Q2) on a 1--5 scale, and whether another word sounds emphasized (Q3).}
\label{fig:ui}
\vspace{-0.5em}
\end{figure}
We conduct objective and subjective evaluations of word-level emphasis control in six languages, measuring how strongly each model emphasizes the target word and how natural the resulting speech is.

The evaluation texts cover six languages (de, en, es, fr, ja, zh), with 35 sentences per language and one annotated target word per sentence, for 210 sentences in total. None of them overlap with the training texts.
We use 8 reference audio clips, 4 male and 4 female.
This yields $210 \times 8 = 1{,}680$ distinct inputs per model.
Since WordVoice supports only en and zh, it covers $8 \times 35 \times 2 = 560$ inputs.
Qwen3-TTS uses 8 of its preset voices (also 4 male and 4 female) and is included only in the objective evaluation, since its voices differ from the reference audio.
To reduce the effect of sampling randomness, we generate 3 outputs with emphasis markers and 3 without for every input.
For the subjective evaluation, one of the 3 marked outputs is selected at random.
The objective metrics are:
\begin{itemize}[leftmargin=*,nosep]
\item $\boldsymbol{\bar{\delta}}$ is the target-word gain $\delta_k$ defined in
  Sec.~\ref{ssec:pairs} averaged over all 3 outputs and all inputs.
\item \textbf{CER}, \textbf{SIM}, and \textbf{NISQA} are defined as in
  Sec.~\ref{ssec:train}; SIM is not reported for Qwen3-TTS, which takes no
  reference audio.
\item \textbf{UTMOS} is the naturalness score of a neural MOS predictor
  \cite{saeki2022utmos}; together with NISQA it serves as an audio quality
  metric.
\end{itemize}
For all metrics other than $\bar{\delta}$, we report the mean over all marked inputs and over all unmarked inputs separately, denoted \emph{emphasis} and \emph{plain}, respectively.

Subjective listening tests remain the more reliable way to evaluate emphasis generation \cite{seshadri2022emphasis,bauer2025explicit,joly2023zerodata,nie2026wordvoice,li2026emorltts}.
We use the same input set as in the objective evaluation and evaluate Ours-Base, Ours-SFT, Ours-DPO, Fish-S2, and WordVoice (en and zh only).
Ours-Base is the pre-trained model without any emphasis marker, serving as a no-emphasis control.
Native speakers of each of the six languages are recruited through Prolific.
Each participant completes 20 trials; the interface and the three questions are shown in Fig.~\ref{fig:ui}: emphasis strength of the highlighted target word (Q1) and overall naturalness (Q2) on a 1--5 scale, and whether any other word sounds emphasized (Q3).
Each participant's 20 trials cover all systems, and between-system comparisons use within-subject paired $t$-tests.

\section{Results}
\label{sec:results}

\subsection{Comparison with existing systems}
\label{ssec:main}

Table~\ref{tab:objective} reports the objective results; $\pm$ denotes the 95\% confidence interval (CI).
Main observations are:
\begin{itemize}[leftmargin=*,nosep]
\item SFT responds to the marker only weakly ($\bar{\delta}=0.069$); DPO reaches $0.136$, slightly above the strongest public system, Fish-S2 (0.129). The explicit acoustic control of WordVoice works to some extent (0.074), while the natural-language instruction of Qwen3-TTS has almost no effect (0.017).
\item The higher CER of the public systems is equally high on plain inputs, i.e., intrinsic to these models rather than caused by the emphasis marker.
\item All systems have similar SIM, UTMOS, and NISQA, suggesting no obvious quality issues.
\end{itemize}

\begin{table}[t]
\caption{Subjective evaluation results: emphasis strength (Q1), naturalness
(Q2), and mis-emphasis rate (Q3). $N$: number of native listeners; mean $\pm$
95\% CI over participants.}
\label{tab:subjective}
\centering
\small
\setlength{\tabcolsep}{4pt}
\begin{tabular}{lcccc}
\toprule
System & $N$ & Q1 $\uparrow$ & Q2 $\uparrow$ & Q3 $\downarrow$\\
\midrule
Ours-Base & 364 & 2.73 $\pm$ 0.08 & 3.45 $\pm$ 0.08 & 42\%\\
Ours-SFT & 364 & 3.15 $\pm$ 0.08 & \textbf{3.52 $\pm$ 0.08} & 34\%\\
Ours-DPO & 364 & \textbf{3.56 $\pm$ 0.07} & 3.47 $\pm$ 0.08 & 32\%\\
Fish-S2 & 364 & 3.25 $\pm$ 0.08 & 2.90 $\pm$ 0.11 & \textbf{29\%}\\
WordVoice (en/zh) & 140 & 3.37 $\pm$ 0.12 & 3.13 $\pm$ 0.15 & 36\%\\
\bottomrule
\end{tabular}
\end{table}

\begin{figure}[t]
\centering
\includegraphics[width=0.95\columnwidth]{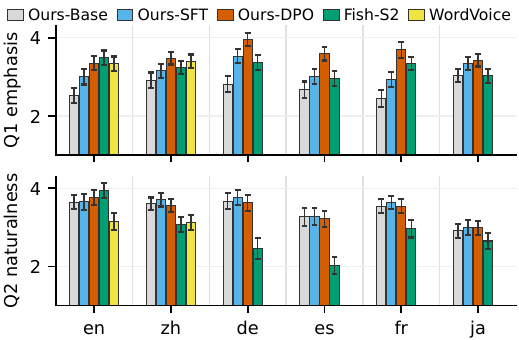}
\vspace{-0.5em}
\caption{Per-language subjective results: Q1 emphasis strength (top) and Q2
naturalness (bottom). WordVoice covers only en/zh. Bars: mean over participants;
error bars: 95\% CI.}
\label{fig:perlang}
\end{figure}

\begin{table}[t]
\caption{Replay strategies, evaluated on the emphasis outputs. SFT $+$10k: SFT trained for 10k more steps; corpus / chosen: replay on real speech / on self-generated chosen samples; LUFS: integrated loudness.}
\label{tab:replay}
\centering
\small
\setlength{\tabcolsep}{2.5pt}
\begin{tabular}{@{}lccccc@{}}
\toprule
Variant & $\bar{\delta}\uparrow$ & CER$\downarrow$ & LUFS & UTMOS$\uparrow$ & NISQA$\uparrow$\\
\midrule
SFT (init) & 0.069\,$\pm$\,0.003 & 0.007 & $-$19.7 & 3.58 & \textbf{4.42}\\
SFT $+$10k & 0.069\,$\pm$\,0.003 & \textbf{0.005} & $-$19.7 & 3.56 & 4.41\\
DPO-chosen & \textbf{0.136\,$\pm$\,0.005} & 0.006 & $-$14.0 & \textbf{3.59} & 3.74\\
DPO-corpus & \textbf{0.136\,$\pm$\,0.005} & 0.008 & $-$19.6 & 3.55 & 4.34\\
\bottomrule
\end{tabular}
\end{table}
Table~\ref{tab:subjective} reports the subjective results and Fig.~\ref{fig:perlang} breaks them down by language. Main observations are:
\begin{itemize}[leftmargin=*,nosep]
\item Emphasis strength (Q1): DPO is rated highest, significantly above Base, SFT, and Fish-S2, and on par with WordVoice on en/zh.
\item Naturalness (Q2): DPO stays at the level of Base and SFT, whereas Fish-S2 and WordVoice are significantly lower. Together with Q1, this shows that DPO strengthens emphasis control with only a few thousand training steps and pairs.
\item Mis-emphasis rate (Q3): DPO is comparable to SFT and slightly worse than Fish-S2.
\item Per language, Fish-S2 beats DPO on en but loses on the others. By checking the Chinese (zh) samples (see the audio samples page), we found that Fish-S2 output carries a noticeable foreign accent. We suspect that Fish-S2 carries over the accent of the English reference audio together with the voice, which would explain why it does well only on en.
\item WordVoice matches DPO in strength but not in naturalness. By checking samples, we found that with fixed pitch/energy/duration settings it merely raises the loudness of the target word, which sounds natural only in some contexts.
\end{itemize}

\subsection{Comparison of replay strategies}
\label{ssec:replay}
Table~\ref{tab:replay} compares the two replay sources of Sec.~\ref{ssec:dpo}---the real SFT corpus (corpus) and the model's own chosen samples (chosen)---with identical initialization, preference pairs, and hyperparameters; an additional row continues SFT for another 10k steps as a control.
Besides the metrics of Table~\ref{tab:objective}, we report the integrated loudness (LUFS); all metrics are computed on the emphasis outputs only.

Training SFT for 10k more steps barely changes any metric, so the DPO gain does not come from the additional updates.
We also tried DPO without any replay, but the denoising errors on both chosen and rejected samples explode after about 2k steps, and the final model can no longer generate intelligible speech.
We suspect that the DPO loss constrains only the difference between the two denoising errors and does not prevent the policy from drifting away from both samples simultaneously.
Replay provides the missing anchor.

The two replay sources reach the same $\bar{\delta}$ but differ in a side effect: with chosen replay, the output becomes 5.7\,dB louder than SFT and NISQA drops from 4.4 to 3.7, whereas corpus replay keeps loudness at the SFT level.
Given this result, we use corpus replay in the objective and subjective evaluations.

\section{Conclusion}
We presented \sys, a simple, efficient, and effective recipe for word-level emphasis control in a multilingual zero-shot diffusion TTS model.
A single off-the-shelf emphasis detector supplies the training signal for both SFT and DPO, so no manual emphasis annotation or task-specific reward design is needed.
DPO substantially strengthens emphasis control with only a few thousand preference pairs and training steps, provided that a replay term anchors the model to real speech.
Objective and subjective evaluations against strong public systems show that our model produces perceptible emphasis while preserving naturalness and overall quality.

\vspace{2pt}\noindent\textbf{Acknowledgments (AI-use disclosure).}
LLM-based coding agents (Anthropic Claude, OpenAI GPT, and Moonshot Kimi models) assisted with the training/evaluation code, produced Fig.~\ref{fig:overview} and Fig.~\ref{fig:perlang} from the authors' design and data, and rephrased and proofread the text of all sections; the authors verified all content.

\bibliographystyle{IEEEbib}
\bibliography{refs}
\end{document}